# "Selfish" algorithm for optimizing the network survivability analysis

Svetlana V. Poroseva

*Mechanical Engineering, MSC01 1150, 1 University of New Mexico, Albuquerque, NM 87131-0001*
1(505)2771493
1(505) 2772761
 poroseva@unm.edu

**Abstract**: In Nature, the primary goal of any network is to survive. This is less obvious for engineering networks (electric power, gas, water, transportation systems etc.) that are expected to operate under normal conditions most of time. As a result, the ability of a network to withstand massive sudden damage caused by adverse events (or *survivability*) has not been among traditional goals in the network design. Reality, however, calls for the adjustment of design priorities. As modern networks develop toward increasing their size, complexity, and integration, the likelihood of adverse events increases too due to technological development, climate change, and activities in the political arena among other factors. Under such circumstances, a network failure has an unprecedented effect on lives and economy. To mitigate the impact of adverse events on the network operability, the survivability analysis must be conducted at the early stage of the network design. Such analysis requires the development of new analytical and computational tools. Computational analysis of the network survivability is the exponential time problem at least. The current paper describes a new algorithm, in which the reduction of the computational complexity is achieved by mapping an initial network topology with multiple sources and sinks onto a set of simpler smaller topologies with multiple sources and a single sink. Steps for further reducing the time and space expenses of computations are also discussed.

# 1 Introduction

In Nature, the primary goal of any network is to survive, that is, if not to avoid, then, withstand life-threatening damage. To be life-threatening, damage should be sudden and either be very precise to destroy vital network elements or be of large scale to eliminate a possibility for a network to recover in a short term. It is rather clear that the latter scenario of massive sudden damage is more likely to be successful.

Surviving as a goal is less obvious in a case of man-made networks, particularly those that are designed for utilitarian purposes (electric power, gas, water systems etc.). These networks are expected to operate under normal conditions most of time. Operational faults associated with such conditions, for example, manufacturing faults, fatigue cracking, and maloperation (Lancaster 2000), are considered to be predictable, random, and repairable within an estimated time. That is, such faults are manageable and, therefore, not an ultimate threat for a network.

Reality, however, changes fast and dramatically. Unprecedented development in science and technology, changes in economical models on global and local scales, the growth of population and the change of its distribution over the planet have brought new challenges for engineers to face. Among the most difficult and costly challenges is the increased likelihood of massive unexpected damage of networks caused by adverse events: natural disasters (hurricanes, earthquakes, floods, wild fires), hostile disruptions (physical destructions, electronic intrusions), man-made errors, and unforeseen combinations of events.

Certainly, some of these threats are not new, but their likelihood is visibly increasing due to climate change, events in political arena, easy public access to information and technologies, and various other factors. Many new technologies along with perspectives of life improvement bring opportunities to misuse them, thus, further contributing in the increased likelihood of adverse events.

Due to increased scale, complexity, and integration of modern networks, the impact of their failure on lives and economy also shows unmatched growth (Lancaster 2000). Some networks, for example, telecommunication and financial systems, have already reached the global level. World economical crises are the best and arguably most unfortunate example to demonstrate how failures in hyper-large networks affect societies worldwide.

The network integration has yet to reach the global level, but it has already been implemented on the country scale. Seven of the Nation's eight critical infrastructures - telecommunications, natural gas and oil, banking and finance, transportation, water supply systems, government services, and emergency services – are directly linked to the electric power infrastructure (President's Commission 1997). Yet, multiple studies (see, e.g., NAERC 2007; U.S. DOE 2002, 2005) indicate that in its current state, the modern electric power infrastructure is not prepared to withstand many forms of large-scale damage. NASA-funded study (CSEISSWE 2008) reported that the loss of electricity would cause "water distribution affected within several hours; perishable foods and medications lost in 12-24 hours; loss of heating/air conditioning, sewage disposal, phone service, fuel re-supply and so on." An example of the impact of the loss of electricity on the society is the biggest blackout in the U.S. history that occurred on August 14, 2003. It affected two countries, left 60 million people in dark, caused multiple deaths, and resulted in estimated 6 to 10 billion economic losses (NextGen ECMIS, 2008). That blackout covered only a part of the entire U.S. electric power infrastructure.

Whereas adverse events in engineering networks cannot be avoided, their scale and impact can and should be mitigated. From the scientific point of view, it can be done by addressing the network ability to withstand massive sudden damage (*survivability*) at the early stage of the network design. This is the ultimate goal of our study.

Many factors contribute to the network survivability (Oliveto 1998; Hill 2005). This is probably one of the reasons why there is no generally accepted definition of survivability. Instead, it varies from one research area to another and from one application to another. More discussion on this topic can be found, for example, in Poroseva et al. (2009), Saleh and Castet (2010). Here, we just specify that the current study concerns with survivability of networks with multiple sources and multiple sinks, where a "source" is a network element generating a quality or service of interest and a "sink" is an element consuming this quality or service. An example of such a network is a power system, with sources and sinks being generators and loads, respectively. In such networks, survivability is perceived as continuation of supply of a quality of interest or service from sources to sinks in the amount sufficient to satisfy their demand in the presence of multiple faults in network elements.

Depending on the origin and evolution of faults, the amount of a quality of interest available to each sink may vary with time. Intuitively, however, the word "survivability" is associated not with a process, but with a final state of a system after all possible damage occurred and before any recovery action can take place. Indeed, one declares "I survived a fire (war, tornado etc.)" only after the event took place, not during its development.

The current study adopts the same approach, that is, only a final steady state of a network is considered after all faults (including cascaded and secondary) have occurred and before any repair has been accomplished. As the origin and the development of faults are not of concern of the current study, the research is applicable to adverse events of any type and intensity in which multiple faults have resulted either due to multiple events (independent or correlated), or due to a single event that caused cascaded/secondary faults in multiple network elements.

In a final steady state, one determines whether a network survived an adverse event by comparing the amount of a quality of interest (service) available in a transformed network with the demand for this quality/service. A mathematical problem to be solved is to establish sources that survived faults, determine their capacity, and verify their connection to sinks. The problem should be solved for all possible combinations of faults.

The necessity to consider all possible combinations of faults comes from the fact that adverse conditions are by nature unexpected in regard of time, space, and scale. That is, it is impossible to predict what combination of faults will come to realization. Traditional engineering approach to identify the worst case scenario and prepare a network for such a case is not applicable to modern networks due to their scale and complexity. Unfortunate proof of this statement is catastrophic events. Their analysis consistently demonstrates that they were caused by an unforeseen combination of events that resulted in an unexpected combination of faults in network elements.

Previously, we suggested (Poroseva et al. 2005) a probabilistic framework for quantifying the network survivability and developed (Poroseva et al. 2009) an efficient computational algorithm for conducting the survivability analysis of

small- to medium-sized networks with multiple sources and a single sink. This approach is applicable, for example, to power grids, when a single sink represents either an isolated industrial load, or multiple commercial and residential loads interconnected into a single distribution system, or a lower-voltage level network.

In larger and more complex networks, the algorithm becomes computationally unfeasible. The main factor is a size of the problem. Indeed, in a network with $M$ elements, the total number of possible final steady states is $2^M$. (In a final steady state, each element can only be in one of the two states: available or faulty). As a simple example of the problem size, the total number of final steady states possible in a network of 50 elements is $10^{15}$. Networks of practical interest may include several thousand of elements or more. From the computational complexity point of view, the problem belongs to the class of exponential time problems (Garey and Johnson 1979; Goldreich 2008) at least. That is, the problem is more complex than more familiar NP-complete problems (see, e.g., Papadimitriou 1994).

The second contributor in the problem complexity is a graph search algorithm that runs for every combination of faults to establish the connection between survived sources and sinks. Presently, this is accomplished by the standard depth-first search traversal algorithm (see, e.g., Cormen at al. 2001). Although actual time required to run this algorithm may vary depending on its application, the theoretical upper bound is linear in the graph size, here, $M$. Other search algorithms are either comparable in performance (e.g., breadth-first search algorithm) or logarithmic in (Goldreich 2008) the graph size.

A rough estimate of the combined contribution of the two factors is $M \cdot 2^M$.

Some ways to reduce the problem complexity were suggested in Poroseva et al. (2009) and Poroseva (2010a). For example, substantial reduction in computational time can be achieved by using an adaptive algorithm to generate final steady states (Poroseva et al. 2009). The adaptive algorithm automatically chooses the deterministic approach where it is feasible and the Monte Carlo technique when computations become expensive. To compare, in a network with the dual bus topology (Doerry 2006) that include 100 sources and one sink, the deterministic approach alone would have to generate and analyze $3.97 \cdot 10^{208}$ final states corresponding to 349 simultaneous faults anywhere in a network. The Monte Carlo technique only generates about one million final states for the same number of faults to conduct the network survivability analysis with the error tolerance of 1E-3. The algorithm parallelization further accelerates the process (Poroseva et al. 2009)

In hierarchical (or multi-layered) networks such as, for example, utility power grids where power is transmitted from the bulk power sources to loads over transmission, subtransmission, and distribution networks operating at different voltage level (Saadat 1998), the problem can be reduced from the analysis of the whole network to the analysis of the survivability of individual network layers. Indeed, survivability of a network as a whole is determined by a network layer with the lowest survivability (the principle of the weakest link in a chain) (Poroseva 2010a). As a scale of an individual network layer is much smaller than a scale of the whole network, the required computational time and space are reduced considerably. If a network has $n$ layers with the $i$-th layer containing the $M_i$ number of elements, then, a rough estimate gives that the problem size is reduced from $2^{\sum_{i=1,\ldots,n} M_i}$ to $\sum_{i=1,\ldots,n} 2^{M_i}$ at least, where $\sum_{i=1,\ldots,n} M_i = M$ and $M_i < M$ for any $i$ when $n > 1$.

The current paper discusses an optimization of a graph search algorithm application. In a new algorithm[1], the problem complexity is reduced by mapping the initial topology of a complex large-scaled network with multiple sources and sinks onto a set of simpler smaller topologies (or *sub-topologies*) with multiple sources and a single sink. As shown in Section 3.2, a search algorithm has to be used only on the set of sub-topologies. Thus, the domain for a graph search algorithm application is reduced drastically and so does the computational time required by the new algorithm.

Philosophy behind this algorithm gives the algorithm its name "selfish" (see discussion in Section 3.2). The algorithm is applicable for evaluating the survivability of networks i) without a recognizable layered structure such as, for example, an individual network layer in a hierarchical network, and ii) in which multiple sinks cannot be represented by a single sink. In combination with the approaches for reducing the computational complexity that were developed in Poroseva et al. (2009) and Poroseva (2010a), the "selfish" algorithm will allow one to conduct the network survivability analysis of real-life networks in manageable time.

---

[1] Initially, the algorithm was presented in the conference paper (Poroseva 2010b) with application to a power system. The current paper, however, is an independent work for the most part.

## 2 General framework of the survivability analysis

### 2.1 Network representation

Networks of practical interest may include variety of different elements, but for the survivability analysis, only four types of elements are of importance: three types of nodes and links that connect the nodes with one another. The three types of nodes are: sources that generate a quality or service of interest, sinks that consume this quality or service, and interconnections through which the quality/service is transported without a change (ideally) in its amount in a direction of a flow of the quality/service. The flow direction through interconnections may vary depending on the network state. Other network elements that are connected in series between any two nodes of the three specified types are represented by a single link between the two nodes. Since a fault in any of the elements connected in series leads to interruption of the quality/service flow through all of them, such simplification of a network is valid. It also reduces the scale of a network to analyze and thus, the computational complexity of the problem.

An example of a network to analyze is shown in Fig. 1. The figure is a modified standard power system diagram of a notional Medium Voltage DC shipboard power system (IEEE 2010) in which all network elements in series (with the exception of sinks, sources, and intersections) are represented by a single link. In the figure, the circles with "-" correspond to sinks (loads), the circles with "+" show sources (generators), and the small dark circles are interconnections.

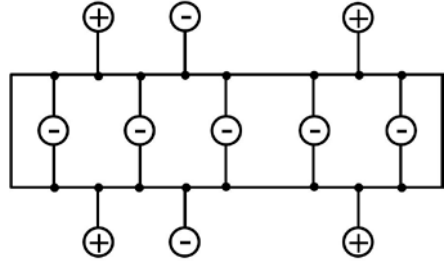

**Fig1** Initial network representation for the survivability analysis

As we are concern with final steady states of a network after damage, faults in network elements are assumed to be failures that cannot be recovered in a short term. Therefore, multiple faults are viewed as simultaneous events; only one fault can occur in a given element. Faulty element is completely unavailable to the flow of a quality of interest or service.

In general, faults can occur in any of network elements. That is, in the network shown in Fig. 1, faults may be in any of four sources, seven sinks, 16 interconnections, and 32 links.

Although the network representation shown in Fig. 1 is appealing due to its visual clarity, the number of possible faulty elements, total 59, is already a computational challenge ($2^{59}$ final steady states in the network). This number can further be reduced for the purposes of the survivability analysis. Indeed, if a source or a sink is connected to the network by a single link, the node and the link are connected in series and therefore, can be represented by a single link. If there are several links connecting a source or a sink to the network, faults in all of them isolate the node from the network, which is equivalent to the fault in the node. Thus, there is no need to consider faults in the nodes separately and they can be removed from the network. Similarly, faults in interconnections can also be removed from consideration.

The result of such simplification of the network in Fig. 1 is shown in Fig. 2. The new representation of the network includes now only 32 links. Faults in these links completely reproduce all possible faults in the network in Fig. 1.

There are three different types of links in the network in Fig. 2: links adjacent to sources (vertical "VT" links), links adjacent to sinks (vertical "VB" links), and links between interconnections (horizontal "H" links). The VB- and VT-links are shown in the figure as arrow-headed links. Four VT-links (VT1-VT4) correspond to the four sources in Fig. 1. The seven sinks in Fig. 1 are represented by twelve VB-links that connect the sinks to the network. Five of the sinks are connected to the network by two links each (VB11 and VB12, VB21 and VB22, VB31 and VB32, VB41 and VB42, and VB51 and VB52). Each of the links VB6 and VB7 connects one sink to the network.

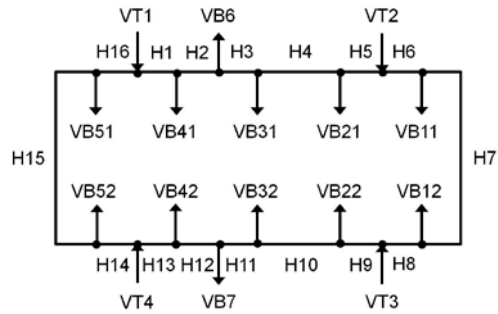

**Fig2** Final network representation for the survivability analysis

To recognize the three types of links mathematically, the VB- and VT-links are assigned the weight. The sign of the weight shows the flow direction through a link: the VT-links have "+" weight and the VB-links have "-" weight. No sign is assigned to the H-links, because the flow direction through such links may vary depending on the network state. A value of the weight assigned to a vertical link is based on its capacity or a maximum amount of a quality of interest (service) it can deliver through in relation to the total demand.

Currently, for the purpose of simplicity only, it is assumed that if multiple VT-links are adjacent to a source, each link is capable to deliver the total amount of a quality of interest (service) generated by the source to the network. In a similar manner, it is assumed that the capacity of each VB-link adjacent to a sink is such that the total demand from the sink for the quality (service) can be satisfied through a single link. The capacity of the H-links allows any required amount of the quality (service) to flow through. These assumptions can be easily adjusted when the analysis is applied to a specific network. Also, additional weights can be added to the links. However, since the current paper concerns with the computational complexity of the survivability analysis, the simplified assumptions serve their purpose.

Notice that the network representation described above puts the emphasis on faults in links instead of faults in nodes. This is an essential difference of our survivability analysis from the traditional reliability/availability analysis. In the latter, the main concern is the availability of the system equipment, whereas the availability of wires (links) is assumed. Under many adverse conditions, however, damage is originated externally and results in physical damage of network elements. From all network elements, links are the most vulnerable to physical damage as they are the least protected due to their extent, protection cost, and exposure to elements, to mention just a few reasons. In some adverse events, such as, for example, blackouts in power grids, links although not destroyed are tripped off, that also make them unavailable for the grid. Thus, with the possible exception of wireless networks and damage initiated and spread within a network (e.g., cyber attacks in communication systems), faults in links are the most expected in adverse events. Since faulty links can separate perfectly functioning sources and sinks from one another, reliability of equipment (availability of nodes) becomes less important factor for the survivability analysis. Moreover, as discussed above, faults in nodes can be represented by faults in adjacent links.

In the mainstream network analysis, the focus is also on the effect of removing nodes rather than links from a network (see, e.g., Newman et al. 2006). After attempting to follow this path, we realized that massive unpredicted damage that can affect any element in a network of arbitrary topology can be more accurately represented by faults in links. Indeed, if a node has multiple adjacent links and only some of them fail, this situation is not easy to describe by removing the node. Moreover, for survivability of a real-life network, it can be of crucial importance to identify what links exactly are faulty from those adjacent to the node. Engineering parameters specific for a particular real-life network (individual link capacity, weight, length, cost, probability of failure etc.) are also easier to assign in the network representation by links.

## 2.2 Fault scenarios

A combination of faults in a final steady state of a network is called a fault scenario. For a given fault scenario, a general problem formulation is to establish

- how many sources remain in operational conditions (survived);
- how many survived sources are connected to sinks;
- whether survived sources connected to sinks can satisfy their demand.

In regard to the demand, a fault scenario can be one of three types depending on a network response it causes: "no response", reconfiguration, and complete failure. In a "no response" scenario, faults do not reduce supply of a quality of interest (service) from sources to sinks. A fault scenario, in which the supply is reduced, but not interrupted, requires the network reconfiguration. That is, some sinks must be intentionally disconnected from the network to balance the demand and the available amount of a quality of interest or service. Such a scenario is called a reconfiguration scenario. If faults completely isolate sinks from sources, a fault scenario is that of the complete network failure.

To determine a fault scenario type, one has to compare the available amount of a quality of interest or service with the demand for this quality (service). Let $Q_D = \sum_{b=1,...,B} Q_{D_b}$ be the demand existing in a network before faults, and $Q_G = \sum_{t=1,...,T} Q_{G_t}$ be the total amount of the quality (service) that can be produced by available sources. Here, $B$ and $T$ are the numbers of sinks and sources, respectively. In Figs. 1-2, $B = 7$ and $T = 4$. Before damage, it is typical that $Q_G \geq Q_D$. If after faults, $Q_G \geq Q_D$, this is a "no response" fault scenario. If $\min[Q_{D_b}] \leq Q_G < Q_D$, this is a reconfiguration scenario. If $Q_G < \min[Q_{D_b}]$, this is a case of the complete network failure.

In a specific network, a threshold to consider the network in a state of complete failure may be higher than the minimum demand from a single sink used here. In such a case, the criteria for reconfiguration and complete failure scenarios should be adjusted accordingly.

Notice that the amount of a quality of interest or service available after damage should be compared with the demand existing in the network before faults occurred. This corresponds to the worst case in a sense of the network survivability, that is, all sinks survived faults and have a request for the quality/service. Thus, this is the best case to test the network ability to withstand damage.

## 2.3 Network response probabilities

In a network with $M$ elements, there are $N = 2^M$ possible fault scenarios. As discussed in Section 1, the only way to prepare a network to adverse events is by analyzing the impact of all fault scenarios on the network operability assuming that they all are equally likely.

The total number $N$ of all fault scenarios is a sum of $S$ "no response" fault scenarios, $R$ reconfiguration scenarios, and $F$ scenarios of the complete network failure: $N = S + R + F$.

A number of fault scenarios leading to a specific network response can be used to determine the response probability $P$ of a network to such scenarios: $P(S) = S/N$ (probability of "no response" scenarios), $P(R) = R/N$ (probability of reconfiguration scenarios), and $P(F) = F/N$ (probability of complete failure).

In a similar manner, one can define the network response probabilities at a given number of faults $m$:

$$P_m(S) = S(m)/N(m), \; P_m(R) = R(m)/N(m), \; P_m(F) = F(m)/N(m), \quad (1)$$

where $\sum_{m=1,...,M} N(m) = N$, $\sum_{m=1,...,M} R(m) = R$, $\sum_{m=1,...,M} S(m) = S$, $\sum_{m=1,...,M} F(m) = F$.

The response probabilities sum to unity.

Such definition of the response probabilities is valid under the assumption that each fault scenario is equally likely. Since damage caused by adverse events is, by definition, unpredictable, this is a justified assumption. However, this is not a requirement of the analysis. If information is available, the probability of each fault scenario can be incorporated into the analysis. Generally though, such information is proprietary and therefore, this case is not considered here.

The three response probabilities completely characterize survivability of a network and can be used to compare the performance of alternative network designs in regard to desirable design features, which are high $P(S)$, low $P(F)$, and a high ratio of $P(S)/P(R)$. The latter is important because higher ratio indicates higher ability of a network to withstand damage without requiring reconfiguration. Under adverse events, reconfiguration itself may not be possible or be too costly as it requires sharing valuable resources such as control and communication with other vital network functions.

The objective of the network survivability analysis is to determine the network response probabilities.

This requires the knowledge of numbers *S*, *R*, and *F* or, equivalently, of numbers $S(m)$, $R(m)$, $F(m)$, and $N(m)$. The total number of fault scenario at a given number of faults can be calculated analytically:

$$N(m) = M\,!/m!(M-m)! \qquad (2)$$

Other numbers can be calculated analytically only in very small and simple networks (Poroseva et al. 2005). As the number of network elements and the network complexity grow, computations become the only choice to determine them.

## 3 Computational analysis

### 3.1 Brute-force algorithm

A brute-force algorithm for calculating the network response probabilities has the structure shown in Fig. 3.

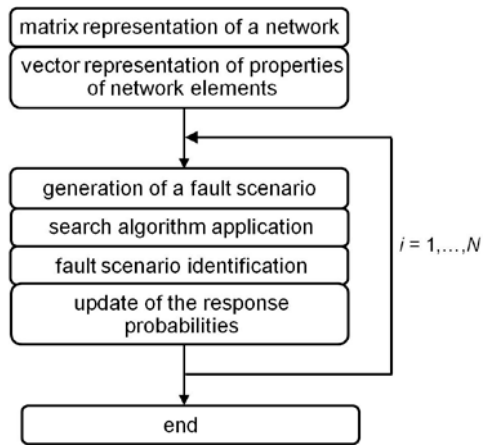

**Fig3** Brute-force algorithm for the survivability analysis

The analysis starts from transforming a network into a structured adjacency matrix or an adjacency list (Poroseva et al. 2009). A standard adjacency matrix is the $M \times M$ symmetric matrix **X** that describes the connectivity of *M* network elements. Here, network elements are the three types of links (see Fig. 2 for an example). If two elements *i* and *j* are connected with one another, we say that $X_{i,j} = 1$ and the two elements are adjacent. If elements *i* and *j* are not directly connected with one another, then, $X_{i,j} = 0$. Each row (column) of **X** is a complete description of all adjacent and non-adjacent elements.

A structured adjacency matrix (list) is a standard one where columns (rows) are ordered. That is, positions $i, j = 1,...,VT$ are reserved for the VT-links, and following positions $i, j = VT+1,...,VT+VB$ are kept for the VB-links. Here, *VT* and *VB* are the numbers of the VT- and VB-links in the network, respectively. The remaining elements ($i, j = VT+VB+1,...,M$) are the H-links. The use of the structured matrix (list) reduces computational time required by a search algorithm to establish the connection between sources and sinks after faults (Poroseva et al. 2009).

The adjacency list organizes adjacent elements into lists or sets. That is, for each network element, there exists the corresponding list that contains only elements adjacent to the element. The use of the adjacency list is particularly beneficial for representing sparse networks such as, for example, power systems, since each set is relatively small. Indeed, if $L_i$ is a size of the adjacency list of the *i*-element, then, one has to store *M* vectors of the size $\max[L_i]$ or less instead of the $M \times M$ matrix. Since typically $\max[L_i] \ll M$, saving in the storage space can be substantial.

Manual generation of the adjacency matrix (list) for complex and large-scale networks is exceedingly time-consuming and error-prone process, with no guarantee of the outcome being correct. Errors left unnoticed will lead to wrong conclusions and ultimately to costly mistakes in the network design and operation. To avoid this situation, a Java application was developed in Neumayr and Poroseva (2011) in order to automatically convert a network diagram into a structured adjacency matrix or list.

Different properties of the network elements can be represented by a set of vectors: one property, one vector. The vector size corresponds to the number of network elements *M*. Currently, only one vector is constructed. This vector represents the link capacity and the flow direction of a quality of interest or service through a link (see discussion in Section 2.1). The vector is updated for every fault scenario to reflect faulty elements. If an element becomes unavailable, the property is set to zero for this element. Both vectors, before and after damage, are used to determine a fault scenario type in the "fault scenario identification" step (Fig. 3). The procedure is described in detail in Poroseva et al. (2009).

The following steps of the algorithm – generation of a fault scenario, search algorithm application, fault scenario identification, and update of the response probabilities – are combined into a single algorithmic block.

Within the block, generating a single fault scenario is a challenge by itself (see discussion in Poroseva et al. 2009). The procedure, however, is not computationally expensive.

Once a fault scenario is generated and survived sources are determined, a search algorithm is applied to establish connectivity between operational sources and sinks. Presently, this is accomplished by the standard depth-first search traversal algorithm (see, e.g., Cormen at al. 2001). Although actual time required to run this algorithm may vary depending on its application, the theoretical upper bound is linear in the graph size, here, $M$.

After the connectivity of sources and sinks in the network with faulty elements has been established, a procedure of identifying the network response is conducted. The procedure does not require much of computational resources. Neither does the updating of the response probabilities.

Notice, that the network response probabilities are re-calculated for each fault scenario. It seems to be not quite consistent with expressions (1). The straightforward use of these expressions turned out to be prone to integer overflow for larger networks. Therefore, a recurrence procedure was suggested in Poroseva et al. (2009) to resolve this problem.

The block steps are repeated until all fault scenarios have been analyzed, that is, $2^M$ times in the deterministic approach. Notable reduction in computational time can be achieved by using the adaptive algorithm ( Poroseva et al. 2009), where the deterministic approach is combined with the Monte Carlo technique. It is difficult, however, to evaluate a computational cost of the application of the adaptive algorithm in an arbitrary network, because the number of generated fault scenarios depends not only on the number of network elements, but also on the required accuracy of calculations. Specifically, at a given number of faults $m$, the algorithm compares $N(m)$ given by (2) and $n(m) \sim \frac{1}{error^2}$, where "error" is a desired accuracy of calculations. If $n(m) \geq N(m)$, then scenarios are generated deterministically. For example, in a network shown in Fig. 2, the application of the adaptive algorithm to calculate the response probabilities with tolerance $error \sim 10^{-4}$ reduces the number of generated fault scenarios from $2^{32} \sim 4.3 \cdot 10^9$ to $2 \cdot 2^{11} + 9 \cdot 10^8 \sim 9 \cdot 10^8$, that is, roughly in about 5 times (or down to $2^{29.7}$).

We will use $2^M$ as the upper bound of computational time required by cycling the block.

Overall, a rough estimate of the required time by the brute-force algorithm is $T_{BF} \sim M \cdot 2^M$. For a network in Fig. 2, it is $T_{BF} \sim 1.4 \cdot 10^{11}$.

To apply the survivability analysis to complex large networks with multiple sources and multiple sinks, additional steps towards reducing computational complexity of the problem should be made.

### 3.2 "Selfish" algorithm

In the mainstream network analysis of the network robustness (see, e.g., Newman et al. 2006), one can say that a network performance is considered from the point of view of an operator, who observes the entire network in all its complexity from aside and for whom the network availability to an individual user is of less importance than the network operability as a whole. Similar viewpoint was also adopted in our previous studies (Poroseva et al. 2005, 2009). However, as we found, the computational complexity of the survivability analysis can only grow on this path with the increase of the network scale and topological complexity.

As often happens in dead-end situations, a solution to a problem is brought by the change of perspective. All engineering networks have been designed not for the sake of their existence, but with the specific purpose to serve the needs of customers. Customers pay for the network design, construction, and operation. If their needs are not satisfied, a network will be of no use. Therefore, the network survivability should also be considered from the customers' perspective which is quite different from the operator's one.

Specifically, a customer, as a matter of fact, is not interested in the network complexity, scale, and its operability as a whole. The only what is of importance is whether the customer has an access to a quality of interest or service and in the amount sufficient to satisfy the customer's demand. Whether other customers have an access to the quality (service) is also out of concern. Such view eliminates from consideration other customers and the multitude of links connecting the customer to available sources. Instead, links are substituted with paths between the customer and a source. Due to obvious reasons, such view on the network operability can be called selfish. This is where the name of the algorithm described below comes from.

From this perspective, the network survivability problem can be re-formulated as to establish the number of paths from a given sink to survived sources in a given fault scenario. Solving the problem for all sinks will give an answer about the network state as a whole in a given fault scenario. Notice that this problem is well suited for parallelization. That is the problem can be solved simultaneously for all sinks, thus, reducing further computational time.

A structure of the numerical algorithm to solve this problem is shown in Fig. 4. The following sub-sections describe the algorithm in detail.

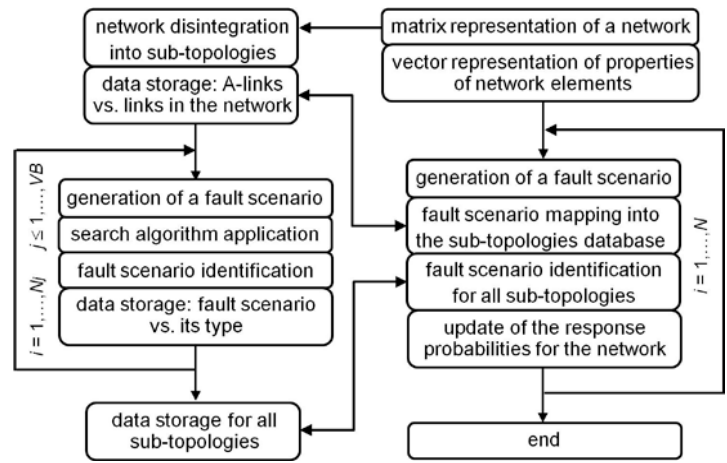

**Fig4** "Selfish" algorithm for the survivability analysis. Here, *VB* is the number of the VB-links in the initial network and *Nj* is the number of fault scenarios in the *j*-th sub-topology

*3.2.1 Network disintegration*

The algorithm starts from disintegrating (or decomposing) the initial network with multiple sources and sinks into a set of simpler networks (or sub-topologies), with each sub-topology containing all sources, but a single sink. This step is performed *before* generating fault scenarios in the entire network.

The network disintegration is achieved by first removing all VB-links belonging to other sinks from the initial network. Indeed, faults in vertical links connecting other sinks to the network cannot interrupt the flow of a quality of interest (service) to the sink under consideration. Then, in each sub-topology, links connected in series are combined into a single link. Since faults in links connected in series are equivalent to a fault in a single link, such procedure is justified for the survivability analysis. Moreover, it reduces the number of links in a sub-topology.

In a general case, the number of sub-topologies is equal to the number of sinks in the initial network. However, one can expect that many sinks "see" the network alike and therefore, the number of sub-topologies to consider will be much less.

The last step in this procedure is the disintegration of sub-topologies with a sink connected to the network by multiple VB-links into sub-topologies with a sink connected to the network by a single VB-link. Not only this step reduces the scale of sub-topologies further, it will also, most likely, reduce their number due to similarity between many of them.

In the worst case scenario, when no similarity between sub-topologies is found, the final number of sub-topologies to consider is equal to the number of the VB-links, that is, *VB*. As only one VB-link is included in a sub-topology, each sub-topology can be uniquely identified by the VB-link.

A general procedure of disintegrating a network with multiple sources and sinks into a set of sub-topologies is shown in Fig. 5.

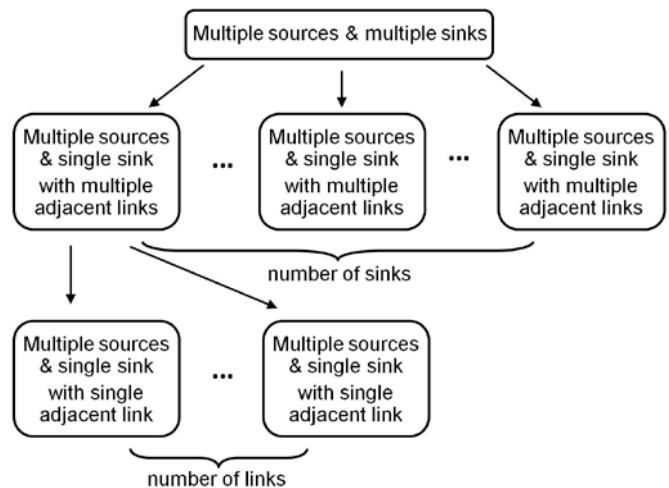

**Fig5** The procedure of disintegrating a network with multiple sources and sinks into a set of sub-topologies with multiple sources and a single sink connected to the network by a single link

Let us consider as an example, the disintegration of the network in Fig. 2. For example, an initial step for the sink connected to the network by vertical links

VB41 and VB42 is shown in Fig. 6a. Figure 6b shows the sub-topology after combining links connected in series into corresponding single links.

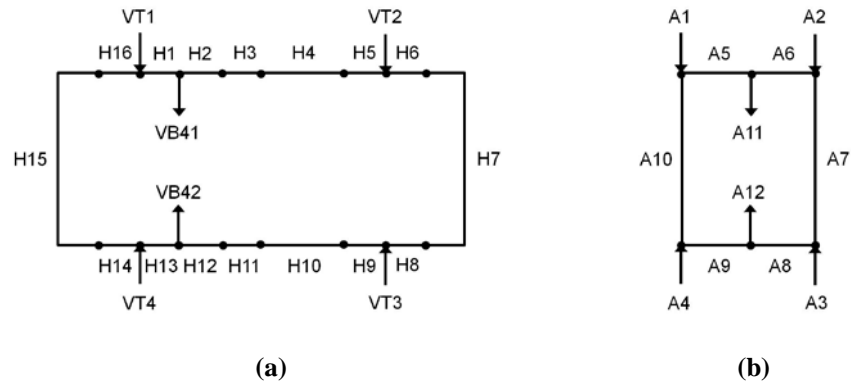

(a)          (b)

**Fig6** A network sub-topology for a single sink: (a) after removing other sinks, (b) after combining links connected in series into a single link

In the figure, the links are re-labeled in such a way that links A1-A4 correspond to links VT1-VT4 and links A11 and A12 correspond to links VB41 and VB42. Links A7 and A10 represent three links each, that is, H6-H8 and H14-H16, respectively. Links A6 and A8 represent four links each: H2-H5 and H9-H12, respectively. There is one-to-one correspondence between links H1 and A5 and H13 and A9. Re-labeling is necessary to simplify the computational analysis as it will be explained later in the section.

Two more sinks "see" the network in Fig. 2 as a sub-topology shown in Fig. 6b. Overall, there are only three unique sub-topologies to consider for the seven sinks in Fig. 2. In addition to the sub-topology in Fig. 6b, these are the sub-topologies shown in Fig. 7.

Each of the three sub-topologies contains 12 or less elements in comparison with the 32 elements of the network in Fig. 2. The A-links in Figs. 6b, 7a, and 7b represent different links in the initial network depending on a sink under consideration.

Further disintegration of the sub-topologies in Figs. 6b and 7b shows that both of them can be reduced to the sub-topology shown in Fig. 7a.

Thus, instead of 12 sub-topologies ($VB = 12$ in the network in Fig. 2), the network can be reduced to a single sub-topology (Fig. 7a) that contains 10 links only.

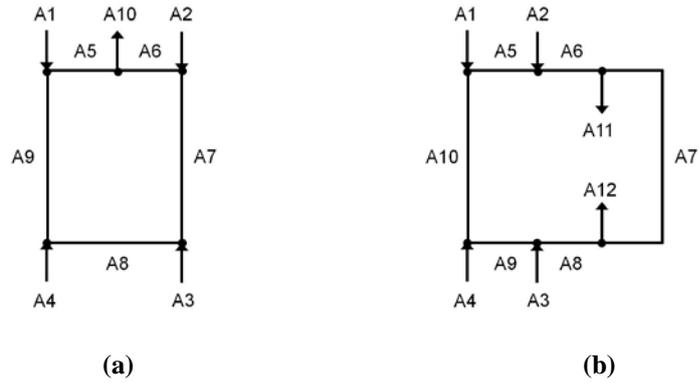

(a)  (b)

**Fig7** Additional two sub-topologies for the network in Fig. 2

The correspondence of the A-links in Fig. 7a to the links in the initial network in Fig. 2 depends on a VB-link that a sub-topology contains. This information is stored in a table form (Table 1). In the table, the A10-link is uniquely reserved to a VB-link. For example, if a sub-topology to be analyzed contains the VB11-link, one finds the "VB11" cell in the A10-column in Table 1. Cells in the row containing "VB11" cell give information on the correspondence of the A-links and the links in Fig. 2 for this sub-topology.

In an engineering network, there is no reason to expect either an excessive number of sub-topologies or that they will be large-scale and complex, at least, not as large and complex as the initial network. That is, tables like Table 1 will not be expensive in a sense of computational space to store data and time to search them.

*3.2.2 Analysis of sub-topologies*

After the initial network was disintegrated and a table similar to Table 1 was generated, the survivability analysis of each sub-topology is conducted separately. That is, in each sub-topology, all possible fault scenarios are generated, a search algorithm is applied, and a type of every fault scenario is determined. If a sink is initially connected to the network by multiple links, fault scenarios are analyzed separately in each sub-topology, where the sink is connected to the network by a single link. A table is generated for each unique sub-topology with information about each fault scenario and its type. Table 2 shows as an example a few rows from the final table for the sub-topology in Fig.7a.

Tables for all sub-topologies constitute a database that is obtained before generating and analyzing fault scenarios in the entire network.

Without taking into account symmetries in the initial network, a rough approximation of the upper bound of computational time required for the analysis of sub-topologies is $\sum_{j=1,...,VB} M_j 2^{M_j}$, where $M_j$ is the number of links in the j-th sub-topology and $2^{M_j}$ is the number of fault scenarios $Nj$ in the sub-topology.

One can show (see Appendix A) that

$$\sum_{j=1,...,VB} M_j 2^{M_j} \leq VB \max[M_j] 2^{\max[M_j]} < M 2^M, \qquad (3)$$

for any $1 < VB \leq M-1$, $M \geq 3$, and $\max[M_j] \leq M-1$. The equality $\sum_{j=1,...,VB} M_j 2^{M_j} = VB \max[M_j] 2^{\max[M_j]}$ holds true when a network is disintegrated into a set of unique sub-topologies that all contain the same number of links ($\max[M_j] = M_j$) and the number of the sub-topologies is exactly VB.

The limits on the values of VB, M, and $\max[M_j]$ are determined by the following. The algorithm can be applied to a network that has at least 2 VB-links (two sinks connected by a single link each to the network). Thus, $VB > 1$. The upper bound on the value of VB comes from the fact that a network should have at least one source (VT-link). The value of $\max[M_j]$ cannot exceed $M-1$ which would be again a case of a network with two VB-links and no possibility to combine horizontal links after removing one of the VB-links. As for the number of network elements M, it cannot be less than three, because it should contain at least two VB-links and one VT-link.

Actual required time for the survivability analysis of sub-topologies is expected to be much less than given by (3). Let us consider the network in Fig. 2 as an example again. As shown previously, only one sub-topology should be analyzed for this network. The total number of faults scenarios in the sub-topology is $2^{10} \sim 10^3$. That is the domain of the application of a graph search algorithm is reduced remarkably in comparison with the one in the brute-force algorithm ($2^{32}$). Thus, $j = 1$ and time required to generate and analyze all fault scenarios in this sub-topology is $10 \cdot 2^{10} \sim 10^4$ that is quite a change from $T_{BF} \sim 1.4 \cdot 10^{11}$.

*3.2.3 Matching fault scenarios in the network and in the sub-topologies*

After the survivability analysis of sub-topologies was conducted and all relevant data stored in a computer memory, the algorithm returns to the generation and analysis of fault scenarios in the network.

The procedure of generating fault scenarios is the same as in the brute-force algorithm (see Section 3.1). However, after a fault scenario was generated, no search algorithm is used to determine the connectivity of survived sources and sinks in the network. Instead, the fault scenario is matched with fault scenarios already existing in the database for the sub-topologies. First, one uses a table similar to Table 1 to find the correspondence of faulty links in the initial network to the A-links in each sub-topology.

Once a fault scenario is identified for every sub-topology, tables similar to Table 2 are used to determine a fault scenario type. Matching procedure is not expensive in time and space as all table data are ordered.

Let us show how the matching procedure works using the network in Fig. 2. For example, faults are generated in the VT1, H1, H2, H5, and H7 links of the network. Using Table 1, one can find that these faults in the network's links correspond to the faults in the A4, A6, and A9 links in the sub-topology in Fig. 7a that contains the VB11 vertical link. For the sub-topology containing the VB12 link the faults are in the A4, A5, and A9 links. For the sub-topology containing either the VB21 link or the VB41, the fault scenario is the same: faulty links are A1, A5, A6, and A7. In a similar manner, one conducts the matching for all rows in Table 1. After that, Table 2 is used to determine a type of the fault scenario for a sub-topology with a given VB-link.

After matching a fault scenario in the network with those in the sub-topologies, the algorithm combines this information i) for each sink connected to the network by multiple VB-links to determine its individual state and ii) for all sinks to determine the state of the entire network (see discussion in Section 2.2). The procedure of updating the network response probabilities is the same as in the brute-force algorithm.

*3.2.4 Time requirements*

The upper bound for computational time required to conduct the survivability analysis of a network using the "selfish" algorithm can be obtained as a sum of time required for the analysis of sub-topologies and for generating all fault scenario, that is, $T_{selfish} \sim \sum_{j=1,...,VB} M_j 2^{M_j} + 2^M \leq VB \max[M_j] 2^{\max[M_j]} + 2^M$. Appendix B shows that indeed

$$T_{selfish} < T_{BF} = M 2^M \qquad (4)$$

for any $1 < VB \leq M-1$, $M > 3$, and $\max[M_j] \leq M-1$.

Actual advantages of the "selfish" algorithm in time can be much higher. For the network in Fig. 2, a rough estimate gives $T_{selfish} \sim 10 \cdot 2^{10} + 2^{32} \sim 2^{32}$ vs. $T_{BF} \sim 32 \cdot 2^{32}$. With increasing the network scale, the difference in running time required by the two algorithms will be higher.

## Conclusions

A computational analysis of the survivability of engineering networks with multiple sources and sinks falls into the category of the exponential time problem at least, due to the fact that the impact of all possible combinations of faults on the network operability should be analyzed. The brute-force algorithm (Poroseva et al. 2009) uses a graph search algorithm to establish the network connectivity in every possible combination of faults. Connectivity of survived sources and sinks determines a network response to each combination of faults. The search procedure increases significantly total computational time required by the brute-force algorithm.

The paper describes a new "selfish" algorithm that reduces the computational complexity of the survivability analysis by substituting a search problem in the brute-force algorithm, which is the application of a search procedure to every combination of faults in the network, with a decision problem (see Goldreich (2008) for more discussion on search and decision problems). In the decision problem, the network response to a given combination of faults is determined by requesting information from a database previously created.

The database is created by mapping the initial topology of a complex large-scaled network with multiple sources and sinks onto a set of simpler smaller sub-topologies with multiple sources and a single sink connected to the network by a single link. The paper shows that a search algorithm has to be used only on this set of sub-topologies. The scale of sub-topologies is expected to be much less than that of the initial network. The number of sub-topologies is also expected to be very small. In the example considered in the paper, the initial network of 32 elements was reduced to a single sub-topology of 10 elements. As a result, computational time required by a search procedure for this example has been reduced from $\sim 10^{11}$ in the brute-force algorithm to $\sim 10^{4}$ in the new algorithm. Thus, the domain of the application of the search procedure is reduced drastically in the new algorithm.

Reduction of the total time required by all steps in the new algorithm has also been proven. In the example considered in the paper, reduction linear in the network size was achieved. Actual time savings are expected to be higher in application to large-scale complex networks.

The algorithm has a broad application. It can be used in designing practically all wired networks with sources and sinks. Application to wireless networks with heterogeneous elements is on-going research and will be reported elsewhere.

## Appendix A

Let us show that indeed $\sum_{j=1,...,VB} M_j 2^{M_j} \leq VB \max[M_j] 2^{\max[M_j]} < M 2^M$. To prove it is to show that $VB \max[M_j] 2^{\max[M_j]} < M 2^M$, when $1 < VB \leq M-1$, $M \geq 3$, and $\max[M_j] \leq M-1$. The values of $VB$ and $\max[M_j]$ are not independent from one another. The relation between them is given by the following expression: $\max[M_j] \leq M - VB + 1$. That gives us

$$VB \max[M_j] 2^{\max[M_j]} \leq VB \cdot (M - VB + 1) 2^{M-VB+1}.$$

The next step is to show that $VB \cdot (M - VB + 1) 2^{M-VB+1} < M 2^M$. Dividing both sides of this expression by $M 2^M$ results in

$$\frac{VB \cdot (M - VB + 1)}{M \cdot 2^{VB-1}} = \frac{VB}{2^{VB-1}} \left(1 - \frac{VB-1}{M}\right) < 1.$$

Substitution of $x = VB - 1$ into this expression gives

$$\frac{x+1}{2^x} \cdot \left(1 - \frac{x}{M}\right) < 1.$$

Function $(x+1)/2^x$ is monotonically decreasing from 1 to $(M-1)/2^{M-2}$ with $x$ increasing from 1 to $M-2$. Because $M$ is never less than 3, the value of this function is 1 or less.

The value of $x$ is always less than $M$, therefore expression $(1 - x/M)$ is always less than 1.

Thus, the product of $(x+1)/2^x$ and $(1 - x/M)$ is always less than 1, that is, indeed

$$\frac{x+1}{2^x} \cdot \left(1 - \frac{x}{M}\right) < 1.$$

# Appendix B

To prove that (4) holds true is to show that

$$T_{selfish} \sim \sum_{j=1,\ldots,VB} M_j 2^{M_j} + 2^M \leq VB \max[M_j] 2^{\max[M_j]} + 2^M < M 2^M.$$

We will use an approach similar to the one in Appendix A. That is, let us first make a substitution of $\max[M_j]$ with $\max[M_j] \leq M - VB + 1$:

$$VB \cdot (M - VB + 1) 2^{M-VB+1} + 2^M < M 2^M.$$

After dividing both sides of this expression by $2^M$

$$\frac{VB \cdot (M - VB + 1)}{2^{VB-1}} + 1 < M$$

and moving 1 from the left side to the right, one has to prove that

$$\frac{VB \cdot (M - VB + 1)}{2^{VB-1}} < M - 1.$$

After two substitutions, such as, $x = VB - 1$ and $a = M - 1$, a final expression to consider takes the following form

$$\frac{(x+1) \cdot (a+1-x)}{2^x} < a.$$

Dividing both sides of this expression by $a$ gives

$$\frac{(x+1)}{2^x}\left(1 + \frac{1-x}{a}\right) < 1. \tag{5}$$

It was shown in Appendix A that function $(x+1)/2^x$ is monotonically decreasing from 1 to $(M-1)/2^{M-2}$ with $x$ increasing from 1 to $M-2$. Because $M$ is never less than 3, the value of this function is 1, when $M = 3$, or less than 1 when $M > 3$.

Let us discuss the behavior of term $(1-x)/a$. This term is equal to zero when $x$ is equal to 1 and becomes negative with $x$ increasing. Therefore, term $(1+(1-x)/a)$ is equal to 1 at $M = 3$ ($VB = 2$) and is less than 1 for all other values of $x$.

Combining the properties of both terms – $(x+1)/2^x$ and $(1+(1-x)/a)$ – one can conclude that the product of both terms is equal to 1 when $M = 3$ or less than 1 for $M > 3$.

Thus, expressions (5) and (4) are satisfied when $M > 3$. Clearly, the case of $M = 3$ (one source with two VB-links adjacent to it) is not of importance in designing engineering networks. Therefore, we conclude that expression (4) holds true for the purposes of the survivability analysis of engineering networks.

# Figure legends

**Fig1** Initial network representation for the survivability analysis

**Fig2** Final network representation for the survivability analysis

**Fig3** Brute-force algorithm for the survivability analysis

**Fig4** "Selfish" algorithm for the survivability analysis. Here, *VB* is the number of the VB-links in the initial network and *Nj* is the number of fault scenarios in the *j*-th sub-topology

**Fig5** The procedure of disintegrating a network with multiple sources and sinks into a set of sub-topologies with multiple sources and a single sink connected to the network by a single link

**Fig6** A network sub-topology for a single sink: (a) after removing other sinks, (b) after combining links connected in series into a single link

**Fig7** Additional two sub-topologies for the network in Fig. 2

# Tables

Table 1 Correspondence of the A-links in the sub-topology in Fig. 5a (the top row) to the links in Fig. 2 for different sinks connected to the network by one of the twelve VB-links. In the sub-topology, the A10-link is uniquely reserved for the VB-links.

| A1 | A2 | A3 | A4 | A5 | A6 | A7 | A8 | A9 | A10 |
|---|---|---|---|---|---|---|---|---|---|
| VT2 | VT3 | VT4 | VT1 | H6 | H7,H8 | H9-H13 | H14-H16 | H1-H5 | VB11 |
| VT2 | VT3 | VT4 | VT1 | H6,H7 | H8 | H9-H13 | H14-H16 | H1-H5 | VB12 |
| VT1 | VT2 | VT3 | VT4 | H1-H4 | H5 | H6-H8 | H9-H13 | H14-H16 | VB21 |
| VT4 | VT3 | VT2 | VT1 | H10-H13 | H9 | H6-H8 | H1-H5 | H14-H16 | VB22 |
| VT1 | VT2 | VT3 | VT4 | H1-H3 | H5 | H6-H8 | H9-H13 | H14-H16 | VB31 |
| VT4 | VT3 | VT2 | VT1 | H11-H13 | H9,H10 | H6-H8 | H1-H5 | H14-H16 | VB32 |
| VT1 | VT2 | VT3 | VT4 | H1 | H2-H5 | H6-H8 | H9-H13 | H14-H16 | VB41 |
| VT4 | VT3 | VT2 | VT1 | H13 | H9-H12 | H6-H8 | H1-H5 | H14-H16 | VB42 |
| VT4 | VT1 | VT2 | VT3 | H14,H15 | H16 | H1-H5 | H6-H8 | H9-H13 | VB51 |
| VT4 | VT1 | VT2 | VT3 | H14 | H15,H16 | H1-H5 | H6-H8 | H9-H13 | VB52 |
| VT1 | VT2 | VT3 | VT4 | H1,H2 | H3-H5 | H6-H8 | H9-H13 | H14-H16 | VB6 |
| VT4 | VT3 | VT2 | VT1 | H12,H13 | H9-H11 | H6-H8 | H1-H5 | H14-H16 | VB7 |

Table 2 An example of the database for fault scenarios and their types in the sub-topology shown in Fig.7a. Total number of fault scenarios is 1024, only three of them are shown. Links inside {…} are those with faults; *S* and *F* stand for "no response" and failure scenarios, respectively. Here, to determine a fault scenario type, it is assumed that a single source can completely satisfy the sink's demand. Generally, identification of a fault scenario type should be based on parameters of a network under consideration.

| Fault scenario | Type |
|---|---|
| {A4, A10} | F |
| {A3, A5, A9} | S |
| {A5, A6} | F |
| ….. | ….. |